\documentclass[preprint,12pt]{aastex}

\shorttitle{Young Stellar Groups Around Herbig
Ae/Be Stars} \shortauthors{Wang and Looney}

\newcommand{\bdstar}{$\rm BD+40\arcdeg~4124$}

\begin{document}

\title{Young Stellar Groups
Around Herbig Ae/Be Stars: A Low-Mass YSO Census}
\author{Shiya Wang and Leslie W. Looney}
\affil{Department of Astronomy, University of Illinois,
1002 W. Green St., Urbana, IL 61801}
\email{swang9@astro.uiuc.edu, lwl@uiuc.edu}

\begin{abstract}
We present NIR and MIR observations of eight embedded young
stellar groups around Herbig Ae/Be stars (HAEBEs) using archived
Spitzer IRAC data and 2MASS data. These young stellar groups are
nearby ($\leq$ 1 kpc) and still embedded within their
molecular clouds. In order to identify the young stellar objects
in our sample, we use the color-color diagram of J - [3.6] vs. Ks
- [4.5]. The Spitzer images of our sample show that the groups
around HAEBEs, spectral types earlier than B8, are usually
associated with bright infrared nebulosity. Within this, there are
normally 10 - 50 young stars distributed close to the HAEBEs ($<$
1 pc). Not only are there young stars around the HAEBEs, there are
also young stellar populations throughout the whole cloud, some
are distributed and some are clumped. The groups around the HAEBEs
are sub-structures of the large young population within the
molecular cloud. The sizes of groups are also comparable with
those sub-structures seen in massive clusters. Young stars in
groups around HAEBEs have generally larger SED slopes compared to
those outside, which suggests that the young stars in groups are
probably younger than the distributed systems. This might imply
that there is usually a higher and more continuous star forming
rate in groups, that the formation of groups initiates later, or
that low mass stars in groups form slower than those outside.
Finally, there is no obvious trend between the SED slopes and the
distance to the HAEBEs for those young stars within the groups.
This suggests that the clustering of young stars dominates over the
effect of massive stars on the low-mass young stars at the scale
of our study.
\end{abstract}

\keywords{stars: formation---stars: pre-main sequence---infrared: stars}

\section{Introduction}

It has been well known that stars form deeply within molecular clouds,
which provide the raw material.
However, how the star formation begins and proceeds within the molecular cloud
is still one of the most fundamental yet unsettled questions in astronomy.
For the past few decades, a picture
of isolated solar-type star formation
has been developed \citep[e.g.][]{Shu1987}.
The basic idea is that
the forming star gathers mass via active infalling and circumstellar
accretion after the onset of gravitational collapse.
Therefore, the star will experience an envelope phase
(Class 0/I protostar) when there exists an optically thick
envelope with heavily infalling matter, a disk phase (Class II
protostar, also known as T-Tauri star, CTT) when the envelope has been mostly
evaporated but a massive disk is still accreting, and then a pre-main
sequence phase (Class III, also known as weak-line T-Tauri star, WTT)
when most material in the disk has disappeared
but a solar-like planetary system may be developing.
Not only a theoretical portrait, these star forming processes
have been extensively suggested by modern infrared observations
of nearby star forming regions.

Nevertheless, there has been increasing evidence that the majority of
stars in the Galaxy do not form in isolation but
in groups or clusters \citep[e.g.][]{Carpenter2000,Lada2003}.
Especially, isotopic studies of the short-lived radionuclide
in the early solar system suggest the contamination from a nearby supernova
\citep[$\sim$ 0.02-1.6 pc distance, e.g.][]{Looney2006a}, 
which suggests that even
the Sun formed in a cluster environment rather than isolation.
But how does a star form and evolve in a cluster environment
\citep[see reviews, e.g.][]{Clarke2000,Elmegreen2000}?
One would expect that
there might be some modification or fundamental difference
of star formation in a cluster from that in isolation.
This should not be too surprising because the dynamical environment, such as
the gas potential, the UV radiation field, encounter cross section of
cluster members, and turbulence performance, dramatically
varies with the evolution of cluster members.

Clusters are characterized by the number of cluster members,
and often separated as groups (N $<$ 100) and clusters (N $>$ 100)
\citep[][]{Adams2001}.
Larger clusters have higher probability to contain more massive
stars, which are able to produce more feedback on the environment.
These clusters, therefore,
increase the ability to detect environmental effects on young cluster
members observationally, but also increase the observational complexity.
On the contrary, small clusters and groups often contain only one or a few
massive stars, so that
the dynamical effects of massive stars may not
be significant or powerful enough to be detected.
However, these small clusters and groups are
the simplest systems that contain massive stars, so
they can be treated as a starting point or basic unit
of studying the environmental effects in clusters.

In this paper, we focus on these small clusters or groups
around one or a few young
intermediate mass stars ($\sim$ 3-20 M$_{\odot}$)
with age $<$ a few Myrs (HAEBEs) \citep[][]{Herbig1960},
in order to diminish the complexity of the environment.
These kinds of groups are typical star formation systems
in the Galaxy.
The purpose of this paper is to investigate the low mass young stellar population
around HAEBEs using Spitzer observations.
In this paper, we will address the following questions:
(1) How are low mass
young stars distributed compared to the HAEBEs; (2) Is there any clustering or
environmental effects; and (3) What are the IR properties of identified
young stars compared to isolated stars.
Furthermore, in order to clarify the definition of cluster and group,
hereafter we use groups for small clusters and groups (our sample) and use
clusters for all multiple systems in general.

\section{Observations}

\subsection{Sample Selection}

We sample eight young stellar groups around HAEBEs
with archived Spitzer IRAC data.
These young stellar groups are still deeply embedded within their
molecular clouds, with spatial scales of a few pc and ages $<$ a few Myrs.
The age criteria is required so that those low mass group members
are old enough to have been born but young enough to be still
going through infalling and accreting phases.
In addition, there is at least one HAEBE star in the group,
so that there will be at least a few tens
of low mass group members.
Moreover, all of them are nearby ($\leq$ 1 kpc),
that increases the sensitivity to
detect lower mass group members as completely as possible.
Table 1 lists the stellar parameters of the most massive star
or stars in groups. Short description of each group
is also summarized as follows:

\begin{itemize}
\item {\bf BD40} contains three HAEBE stars, \bdstar,
V1686 Cyg, and the southern star of V1318 Cyg binary, with age
$\sim$ 1 Myr at 1 kpc. Molecular clumps have been detected centered
on the southern star of V1318 Cyg, which is therefore believed
to be the youngest and most actively forming star \citep[e.g.][]{Looney2006b}.
Moreover, more than 80$\%$ of the group members show near 
infrared excess \citep[][]{Hillenbrand1995},
which implies the youthfulness of this group. The fact that both
massive and low mass stars form simultaneously within such a small
scale provides us an excellent environment
to study the star formation in the cluster.

\item {\bf NGC 7129} is a reflection nebula that contains three Herbig Be stars,
LkH$\alpha$ 234, $\rm BD+65\arcdeg~1637$, and $\rm BD+65\arcdeg~1638$.
A number of young stars have been found in this system, along with
molecular outflows \citep[][]{Edwards1983,Fuente2001}
and dense molecular clumps centered on LkH$\alpha$ 234 \citep[][]{Wang2007}.
It has been suggested LkH$\alpha$ 234
is the youngest among the three Be stars \citep[][]{Hillenbrand1992}.
Several authors have published
the Spitzer photometric and spectroscopic results for this system
\citep[e.g.][]{Megeath2004,Morris2004,Gutermuth2004}.
It is included as one of our sources for completeness.

\item {\bf MWC 297} is an O9e star at 450 pc
\citep[][]{Hillenbrand1992}. B1.5V at 250 pc is also suggested by
other authors \citep[e.g.][]{Drew1997}. We use the former in this
paper. Dense molecular emission, such as $^{13}$CO and C$^{18}$O
\citep[][]{Fuente2002,Ridge2003}, and molecular outflows
\citep[e.g.][]{Drew1997} have been detected around MWC 297, which
suggests an active star forming environment.

\item {\bf VY Mon} is a B8e star with age $\sim$ 0.1 Myrs,
surrounded by $\sim$ 25 stars \citep[][]{Testi1998}.
$^{13}$CO and C$^{18}$O observations show that the
dense molecular gas is distributed
along an extended ridge around VY Mon \citep[][]{Ridge2003}.
Two 10 $\mu$m companions were detected \citep[][]{Habart2003}.

\item {\bf VV Ser} is a B9e star with age $\sim$ 0.6 Myrs,
surrounded by $\sim$ 24 stars \citep[][]{Testi1998}.
$^{13}$CO and C$^{18}$O observations show that the
dense molecular gas is extended
and VV Ser is located within a cavity of this dense gas
\citep[][]{Ridge2003}.

\item {\bf HD 97048} is an A0e star surrounded by a few stars. It
is the oldest in our sample.

\item {\bf BD46} system is embedded within a molecular cloud
that includes the Herbig A0e star, {$\rm BD+46\arcdeg~3471$}, and
the B0 star, {$\rm BD+46\arcdeg~3474$}.
$\rm BD+46\arcdeg~3474$ illuminates an emission nebula,
IC 5146, that contains more than 100 stars, including
four other late B stars; while
$\sim$ 3.5 pc away, $\rm BD+46\arcdeg~3471$ illuminates another emission nebula
including more than 10 stars with an average age of 0.2 Myrs \citep[][]{Herbig2002}.
How two clusters with such different sizes form closely in the same cloud is still unsolved
and it also makes this system interesting for the study of star formation.

\item {\bf V921 Sco} is an Be star away from the HII region, GAL
343.49-00.03, at the distance of $\sim$ 2 pc. This HII region
contains the IR source, IRAS 16558-4228, and several radio
sources. Four 10 $\mu$m companions around V921 Sco were detected
\citep[][]{Habart2003}.

\end{itemize}

\subsection{Infrared Observations}

In this paper, we study the infrared emission of young stars.
Young stars are normally detected at infrared bands because
their circumstellar dust re-emits the stellar light into
the infrared regime and enhances the infrared luminosity (so called
infrared excess).

We use data from the 2MASS \citep[][]{Skrutskie2006} point source catalog
and the Spitzer Archive
to obtain the photometry at near- (2MASS: J, H, and Ks) and
mid- (Spitzer IRAC: [3.6], [4.5], [5.8], and [8.0])
\citep[][]{Fazio2004} infrared bands.
Table 2 summarizes the observational details of our sources from
the Spitzer Archive.
The source position and photometry of JHKs were obtained directly from the
2MASS point source catalog; the
point source detection and photometry of IRAC data, which are
PBCD in the Spitzer Archive,
were accomplished using IRAF APPHOT package (daofind and phot tasks).
Only sources above 5 $\sigma$ are considered. We use an aperture radius
of 3 pixel (3.6$\arcsec$) and sky annulus from 3 (3.6$\arcsec$) to 7 (8.4$\arcsec$) pixels
for all point sources. We use the zero magnitude of
21.108, 20.635, 20.1667, and 19.5 for IRAC [3.6], [4.5], [5.8],
and [8.0], respectively \citep[][]{Fazio2004}.
The uncertainty of IRAC magnitudes is less than 0.05 mag
for each channel of all sources.
The detection limit at J band, J $=$ 17, is applied, which corresponds to
the 0.2 mag uncertainty.
Table 3 list the detection limit at each band of the identified YSO candidates
in this study. The last column in Table 3 is the spectral type whose
color is the detection limit at [4.5] using Figure 1(b) (more details about Figure 1
in Sec. 3.1).

\section{Identifying Protostars}

Recent studies in nearby embedded clusters
have shown that photometry of
Spitzer observations is efficient to
identify young stars, especially to distinguish
between stars with disks and envelopes
\citep[e.g.][]{Allen2004,Megeath2004,Allen2006}.
For example, \citet{Hartmann2005} plots Taurus young stars
with known classes (i.e. WTTs, CTTs, and Class I protostars),
and shows that all of them are well
separated in the color-color diagram, [3.6] - [4.5] vs.
[5.8] - [8.0].
In addition,
the color-color diagrams that combine IRAC and MIPS bands,
for example [3.6] - [8.0] vs. [8.0] - [24] and
[3.6] - [5.8] vs [8.0] - [24],
are also efficiently used to identify young stars,
especially for highly embedded stars and stars
with significant inner holes
\citep[e.g.][]{Rho2006,Lada2006}.
Indeed, unlike the near infrared (e.g. JHK) emission
from young stars, the
IRAC ([3.6], [4.5], [5.8], and [8.0])
and MIPS [24] channels
contain a significant fraction of emission
from circumstellar material
(disk or envelope) over the stellar photosphere,
so these channels are more sensitive to
probe the circumstellar characteristics than the near infrared.

However, not all sources can be simultaneously detected
at all bands. This is because the [3.6] and [4.5]
are much more sensitive than others \citep[][]{Fazio2004}.
On the other hand, [5.8] and [8.0] contain strong PAH emission
that easily confuse the detection and photometry of point sources.
Since the archived data of our sources are not
deep and the majority of the young star population in
these groups
are low mass, photometry using only the IRAC bands is not reliable
to identify young stars.
Therefore,
due to the intrinsic difficulties of nebulosity confusion and sensitivity
at [5.8] and [8.0],
only the [3.6] and [4.5] IRAC bands are used here
and combined with J, H, and Ks to identify young stellar objects.
Indeed, the J to [4.5] colors
have been used in some studies of embedded clusters
\citep[e.g.][]{Gutermuth2004}.
In order to best
separate between Class I/II protostars and Class III/normal stars
in our sample with only J, H, Ks, [3.6], and [4.5],
we examine the color-color diagrams of one of the most
well studied and nearby young clusters IC 348 using the published infrared
data \citep[][]{Luhman2003,Lada2006},
and apply the results to our sources.

\subsection{Case Study of IC 348}

IC 348 is a partially embedded young cluster with age $\sim$ 2 Myrs at
260 pc, which contains $\sim$ 300 cluster members with spectral
types from early B down to M \citep[][]{Muench2003}.
It is a cluster with an
age and number of cluster members larger
than our sample.
We simply assume that the young stars in IC 348 are generally similar
to those in our sample, so that we can compare techniques to identify
young stars.
Published NIR and Spitzer data of IC 348 are used, including
the colors at J, H, Ks \citep[][]{Luhman2003}, [3.6]
and [4.5], SED-derived A$_V$, and classes of protostars \citep[][]{Lada2006}.
In order to obtain the intrinsic colors,
the interstellar extinctions are corrected by applying
the extinction laws from \citet{Rieke1985}
for J, H, and Ks
and \citet{Indebetouw2005} for the IRAC bands, which corresponds to
R$_V = 5.5$.
Moreover,
in the paper of \citet{Lada2006}, stars are classified as
stars with pure stellar photospheres (star, hereafter called
normal star), optically thick disks (thick),
and anemic disks (anemic). Those stars with optically thick disks are
most likely CTTs that have primordial disks
while those with anemic disks
are more like transition ones with mostly depleted disks.
Here we only include stars with thick disks as young stars,
in order to identify the primordial disk populations.

Figure 1 plots the absolute magnitude at [4.5] of normal and thick disk stars
by applying the extinction and distance corrections.
Using this figure, we can estimated the range of spectral types of
identified young stars at an assumed distance.
Figure 2 plots the dereddened color-color diagrams of
H - [3.6] vs. [3.6] - [4.5] and Ks - [3.6] vs. [3.6] - [4.5] for IC 348,
which are the methods often suggested by
other authors \citep[e.g.][]{Hartmann2005,Allen2006}.
For example, \citet{Hartmann2005} showed that Taurus
Class I/II protostars are clearly distinguished from
Class III sources in both diagrams, especially in the diagram of
H - [3.6] vs. [3.6] - [4.5].
In Figure 2, normal and thick disk stars in IC 348 are
indeed distributed as two groups in the color-color diagrams.
However, normal stars will mix with thick disk stars
when they are highly extincted; extinction is usually
significant in embedded clusters.
In addition, the solid lines in Figure 2,
which were arrows drawn in the Taurus study \citep[Figure 3 in][]{Hartmann2005},
can not separate thick disk stars
from normal stars in IC 348 as well as those in Taurus.

Figure 3(a) and 3(b) plot two dereddened color-color J to [4.5]
diagrams for normal and thick disk stars in IC 348.
They are the best color-color diagrams to separate
normal and thick disk stars.
We derive lines that divide the normal and
thick disk stars,

\begin{equation}
J-[3.6] = \frac{E(J-[3.6])}{E(Ks-[4.5])} \times (Ks-[4.5]) - 0.36
\end{equation}
(solid line in Figure 3(a) and 3(c)) and

\begin{equation}
H-[3.6] = \frac{E(H-[3.6])}{E(Ks-[4.5])} \times (Ks-[4.5]) - 0.2,
\end{equation}
(solid line in Figure 3(b) and 3(d)),
which are both along the direction of extinction with
$E(J-[3.6]) = 0.219 \times A_V$, $E(H-[3.6]) = 0.112 \times A_V$,
and $E(Ks-[4.5]) = 0.064 \times A_V$
\citep[][]{Rieke1985,Indebetouw2005}.
Because Equation (1) and (2) are along the direction of extinction,
normal and thick disk stars can be still separated no matter how large
the extinction.
We also compare these two diagrams to Taurus' results
\citep[][]{Hartmann2005}.
Figure 3(c) and 3(d) plot the colors of Class I-III protostars in Taurus
in the same color-color diagrams as 3(a) and 3(b).
They show that Equation (1) and (2) can also
discriminate most stars with circumstellar material
(CTTs and Class I) from stars without disks (WTTs).
According to the results of IC 348 and Taurus, we can conclude that
at least 90$\%$ of young stars will be identified using
Equation (1) and (2), and at most 10$\%$ of them are
normal star contamination.
In this study, we use Equation (1) and the color-color diagram of
J - [3.6] vs. Ks - [4.5].

Not only can the thick disk stars be identified in Figure 3(a) and 3(b),
they are also distributed within narrow bands.
These are also found for CTTs in the color-color diagrams
of J - H vs. H - K and H - K vs. K - L, so called CTTs loci
\citep[][]{Meyer1997}.
Likewise, we can obtain two loci (dashed lines in Figure 3)
via least-square fittings,

\begin{equation}
J-[3.6] = (1.01 \pm 0.04) \times (Ks-[4.5]) + 0.79 \pm 0.04;
\end{equation}

\begin{equation}
H-[3.6] = (0.95 \pm 0.04) \times (Ks-[4.5]) + 0.12 \pm 0.04,
\end{equation}
which are called YSO loci here. Although it does not necessarily mean that
all young stars will locate along the YSO loci intrinsically,
they can be useful to estimate the interstellar extinction of young stars.
With estimated extinctions, we can get an average extinction, that
help us understand the cloud conditions of our sample.

In summary, after examining the intrinsic colors of young stars in IC 348
and comparing them to the results in Taurus,
we decide to use the color-color diagram of J - [3.6] vs. Ks - [4.5]
as the tool of identifying young stars.
All stars with colors
redder than Equation (1) in this color-color diagram
are young stellar candidates (hereafter called YSO candidates).
This method is useful for those low mass groups similar to Taurus
and intermediate-sized clusters similar to IC 348,
when using J to [4.5] colors only.
Methods of identifying young stars and YSO loci of young stars in different
environments may be different due to the environmental impacts on
the young stars.
Therefore, in the future, this should be tested in other
environments, such as massive clusters.

\section{Spitzer Images and YSO Candidates}

\subsection{Dust Morphology}

RGB color composite images of all sources are plotted in Figure 4,
where [3.6], [4.5], and [8.0] are displayed as blue, green, and
red, respectively. This figure shows the distribution of dust in
red, which contributes significantly at [8.0]. The emission comes
from thermal dust continuum illuminated by the massive HAEBEs and
the PAHs, that often trace the star forming regions. From this
figure, we can see that all but two of our sources show bright
diffuse emission in channel [8.0]. However, the most massive star
in the system, the HAEBE, is not always at the center of the dust.
For example, the BD40 system is in a dust-rich environment and
located right at the intersection of two branches of dust with
different position angles. This is consistent with the two systemic
velocities of dense gas that suggests that the
high star forming activity is due to the collision of clouds.
\citep[][]{Looney2006b}. Unlike MWC 297, which has a basically
centered symmetric dust nebula, the VY Mon and NGC 7129 systems
are located at the edge of dust structures. These one-side
structures may result from the disruption of the natal cloud by
the massive star in one direction. However, we can not exclude the
possibility that massive stars may not primordially form in the
center of a dense cloud. Two of our systems, HD 97048 and VV Ser,
show very little dust emission. This might be because they are
both at the lower mass end of HAEBE stars (3.4 and 3.3 M$\odot$,
respectively)-- not energetic enough to illuminate their
environments.

Interestingly, the bright dust structures in two systems, BD46 and
V921 Sco, are actually not associated with the HAEBEs, {$\rm
BD+46\arcdeg~3471$} and V921 Sco, respectively. In the BD46
system, a symmetric morphology of dust with radius $\sim$ 2 pc is
centered on the B0 star, {$\rm BD+46\arcdeg~3474$}, while no
obvious dust is detected around the A0e star, {$\rm
BD+46\arcdeg~3471$}. Again, this might be because {$\rm
BD+46\arcdeg~3471$} is not massive enough, as it is an A0e star.
On the other hand, in the V921 Sco system, there are multiple dust
accumulations. The brightest dust feature is centered on the HII
region, GAL 343.49-00.03, with radius $\sim$ 1 pc and $\sim$ 2 pc
away from the B0e star V921 Sco. The Spitzer image also shows
there is a large group of optically invisible stars deeply
embedded within this HII region, that might suggest it is the most
active star forming region in the area. Two other small dust
clumps with radius $<$ 0.5 pc are centered on two IRAS sources,
IRAS 16566-4229 and IRAS 16570-4227, $\sim$ 2 and 2.5 pc away from
the HII region, respectively. As shown in Fig. 4, there is also
some dust around the B0e star, V921 Sco. How the small groups of
stars around {$\rm BD+46\arcdeg~3471$} and V921 Sco form so
closely to the large clusters around {$\rm BD+46\arcdeg~3474$}
and the HII region is interesting and still unknown. For the BD46
system, the {$\rm BD+46\arcdeg~3471$} group might form and
escape while the gas within the {$\rm BD+46\arcdeg~3474$}
cluster was dispersing, because {$\rm BD+46\arcdeg~3471$} seems
to be isolated without any diffuse dust connected to the {$\rm
BD+46\arcdeg~3474$} nebula. However, for the V921 Sco system,
the V921 Sco group might form due to the collision of turbulent
clouds near the actively forming cluster (the HII region), because
both of them are surrounded by a large scale diffuse dust.

Table 4 lists the estimated spatial radial scales for the dust
around the massive stars, defined as $R_d$. We briefly conclude
that groups around the HAEBEs whose spectral types are earlier
than B8 are usually associated with bright dust and PAH emissions,
although the formation mechanisms of our sample may be varied due
to their different morphologies.

\subsection{Color-Color Diagrams}

Figure 5 shows the color-color diagrams of J - [3.6] vs. Ks - [4.5] for all sources.
The solid and dashed line are Equation (1) and (3), respectively.
Stars redder (rightward) than the solid lines
are characterized as YSO candidates;
while those YSO candidates above the dashed lines are most likely thick disk
stars (CTTs). From this figure, we can see that most of YSO candidates are indeed
CTT-like. Those YSO candidates below the dashed lines and with large
Ks - [4.5] may be young A or F stars,
similar to those in the color-color diagram of J - H vs. H - K
in Figure 15 of \citet{Hillenbrand1992}.

Table 4 also summarizes the total number of identified YSO
candidates for each source in the whole data field and inside the
group, respectively. We simply define that the stars within the
dust region ($R_d$, estimated in the Sec. 4.1) are inside the
group. \citet{Adams2006} derived the relationship between the
number of cluster members ($N$) and radius of the cluster ($R$),
$R=\sqrt{3}$ pc $(N/300)^{1/2}$. Based on this equation, we
calculate the group radius ($R$ in Table 4) assuming that the
entire population consists of our YSO candidates. The result that
$R$ is comparable or less than $R_d$ (see Table 4) shows that the
number of detected YSO candidates in groups is actually close to
or less than the number of group members. Based on the number of
YSO candidates, our sample can be sorted into three divisions: (1)
$\sim$ 10 YSO candidates for HAEBEs later than B9: {$\rm
BD+46\arcdeg~3471$} and HD 97048; (2) 20-50 for single HAEBEs
earlier than B9: MWC 297, VY Mon, VV Ser, and V 921 Sco; (3) $>$
50 for multiple HAEBEs or B stars: BD40, NGC 7129, and {$\rm
BD+46\arcdeg~3474$}. The HII region in the V921 Sco system is
not included because it is actually not a HAEBE group. This
categorization implies a positive trend between the mass and
number of massive stars and the number of young stars in groups,
just as suggested by other authors
\citep[e.g.][]{Hillenbrand1995,Palla1995}. Despite the
categorization, we can simply conclude that there are normally 10
- 50 young stars distributed centered at or close to the HAEBEs.

Moreover, according to Figure 5, the interstellar extinction for
each CTT-like YSO candidate can be estimated using the offset from
the YSO locus (Equation (3)). Except the V921 Sco system, which
shows a relatively high extinction (A$_V$ $\sim$ 10 - 20 in
average), the majority of the extinction is $<$ 10 (peak at $\sim$
5) for YSO candidates in our sample.

Figure 6 plots those YSO candidates in our sample with the
identification of all four IRAC bands in the color-color diagram
of [3.6] - [4.5] vs. [5.8] - [8.0]. We use the classification
scheme proposed in \citet{Megeath2004} to characterize these YSO
candidates: 10 $\%$, 3 $\%$, 80 $\%$, 1 $\%$, and 6 $\%$ of them
are Class 0/I, Class I/II, Class II, reddened Class II, and normal
stars, respectively. This is consistent with the conclusion that
more than 90 $\%$ of YSO candidates are most likely young stars,
from the case study of IC 348.

There is the possibility of contamination from background
galaxies in our YSO candidate sample.
\citet{Harvey2006} use the Spitzer
Wide-Area Infrared Extragalactic Survey (SWIRE)
Elais N1 data \cite[][]{Surace2004} to determine the extragalactic
contaminates; they find that $\sim$ 25 of the 591 sources (assuming they are all
extragalactic)
within 0.89 degree$^2$ meet their YSO selection criteria.
In other words, on average $\sim$ 25 extragalactic sources are
misidentified as YSO candidates within 0.89 degree$^2$.
Although our YSO selection criteria are different (see Section 3), we can compare
our results numerically,
as most of the sources are foreground or background sources.
For example, in our groups, $\sim$ 5 - 15 $\%$ of identified point sources are
YSO candidates, only slightly larger than the $\sim$ 5 $\%$ in
\citet{Harvey2006}.
This is not surprising as we expect to have more
YSOs in our sample of pointed star formation observations compared to the
large-scale survey.
In other words, the YSO selection criteria are consistent with each other.
Moreover, the detection limit at [8.0] in our sample, $<$ 14.0 (Table 3),
is higher than that of the 25 extragalactic sources.
Therefore, we adopt their value of 25 extragalactic sources in 0.89 degree$^2$ as
an upper limit
to extragalactic contamination of our YSO candidates.
For our sources, with small fields of view, it is unlikely
that there will be more than a few
extragalactic sources in each of our fields.
Therefore, we conclude that the
contamination from background galaxies should be negligible.

\subsection{Spatial Distribution of YSO Candidates}

One of the most basic questions when studying star formation in
clusters is the location of the young stellar population. Two
elements determine the distribution of young stars in the cluster,
the initial cloud condition and the interaction of cluster
members. Clusters form through the fragmentation of the molecular
cloud, which is determined by the initial conditions, such as the
gravitational potential, velocity field of the gas, magnetic
field, and turbulence \citep[see review,][]{Meyer2000}.
However, after the onset of
individual collapse in the cluster, the forming stars will
continue being affected by the dynamical evolution of the cluster, for
example, the removal of gas, feedback of stars, and collision
between cluster members. With the continuous modification of the
dynamical state in the cluster, the distribution of stars might
change dramatically. \citet{Adams2006} modelled the early
evolution (within 10 Myr) of clusters with 100 - 1000 members with
respect to the boundness, cluster size, virial ratio, and number
of cluster members with time. Their results showed that the gas
dispersal plays a primary role in the evolution of these
parameters. For example, in their Figure 3, the fraction of bounded
cluster members is decreased from 1 to 0.2 within 5 Myrs after the
gas dispersed for the virial initial condition. Although this
depends on the real gas dispersal processes dramatically, it
implies that there is a significant fraction of cluster members
that will escape the cluster eventually; there should be a
significant fraction of isolated stars actually formed in
clusters. Moreover, \citet{Allen2006} studied several embedded
clusters from the Spitzer survey and showed that protostars tend
to distribute in elongated morphologies around massive stars,
along with some sub-clustered structures. They suggested that the
distribution of protostars may still appear as the primordial
structure of the natal cloud. Indeed, studying the spatial
distribution of young stars in clusters will allow us to assess
not only the cluster formation but its history.

Figure 7 plots the number density of YSO candidates vs. the distance to
massive stars. For example, in the BD46 system, it is centered on
{$\rm BD+46\arcdeg~3471$}, instead of {$\rm BD+46\arcdeg~3474$}.
The number density defines the number of YSO candidates per
surface area (number pc$^{-2}$).
From this figure, we can see that there are indeed groups of YSO candidates
around the massive stars as expected, but also a young stellar
population throughout the whole field outside the group.
Those YSO candidates outside the group are mostly
distributed uniformly. However, there are some clumps of young stars
outside the groups around HAEBEs, for
example, MWC 297 and BD46.
The young stellar group around HAEBEs, the clumps of young stars
outside the group, and the overall distributed young population
may be formed as a part of sub-clustered structures of an dispersed cluster.
The fact that there exists a young population outside the group is
reasonable as the group is actually
embedded in a large-scale molecular cloud.
The Spitzer observations of several young stellar clusters,
including NGC 7129, show a
similar picture \citep[][]{Megeath2004,Gutermuth2004} --
a significant fraction of identified young stars
distributed outside the cluster core.
Their results also show that these young stars outside the cluster are
actually well associated with the large scale dense molecular gas.
They also suggest that the cluster forming environment
plays an important role in determining the morphology of the young stellar
population.

From Figure 7, we also can estimate the spatial size of the young
stellar group around HAEBEs ($R_n$ in Table 4) and the clumps of
young stars outside the group. These groups and clumps have the
sizes $\sim$ 0.5 - 1 pc, which are similar to the sizes of those
sub-cluster structures in massive clusters \citep[][]{Allen2006}.
This further supports the suggestion in Section 1 that these
groups can be treated as basic units of studying the star
formation in clusters. In addition, Table 4 shows $R_d$ and $R_n$
are basically consistent, as $R_d$ shows the range of the dust
distribution and $R_n$ probes the spatial clustering of the YSO
candidates.

In summary, our results suggest that the groups around HAEBEs
usually contain 10 - 50 young stars within the dust emission and
form relatively isolated groups separated from the other regions
in the large scale molecular cloud. The HAEBEs are normally
located near the center of the group of young stars. These groups
can be treated as sub-structures of the large scale young stellar
population within the molecular cloud, as the clumpness of YSOs
throughout the whole fields are seen. Each clump and group can
also be compared to those sub-cluster structures found in massive
clusters.

\section{SED Slopes vs. Distance to the HAEBEs}

The slope of the spectral energy distribution (SED) is often
used as the parameter indicating the youthfulness of a young star
\citep[e.g.][]{Wilking1989}.
The SED slopes of the YSO candidates in our sample
are also derived, using Ks, [3.6], and [4.5].
However, the SED slope is modified by the interstellar extinction.
We derive the SED slopes of YSO candidates in the BD40 system
with A$_V$ $=$ 0, 5, 10, 15, and 20, and conclude that
the SED slope will change $\sim$ $+$0.4 per increasing
A$_V$ $=$ 5. This provides an uncertainty to
the SED slope derivation, along with the estimated A$_V$ from
Figure 5, for each of our sample.

\subsection{Environmental Impact from Massive Star and Clustering}

Not only are the position and mass of young stars affected by
either the massive star or clustering, but the evolutionary stages
of the clustering stars may also be affected. During the formation
of massive stars, the molecular gas is actively disrupted by their
strong stellar winds and UV radiation fields. These two factors
might effectively alter the formation and evolution of
protoplanetary disks, too. Studies on how EUV and FUV photons
truncate the circumstellar disk and affect the formation of
planets are increasing, especially theoretically. It is
predicted that the mass loss decreases with the FUV flux, thus the
distance to the massive star, so that the disk size would increase
with distance to the massive star within the FUV dominated
regime \citep[][]{Storzer1999}. In addition, alternative models
about how close encounters of cluster members affect the size and
mass loss of circumstellar disks are also studied.
\citet{Pfalzner2006} suggested that the mass loss on the disks
decreases with the distance from the cluster center due to the
star-disk encounter, and high and low mass stars are affected more
seriously than intermediate mass stars by the cluster environment.
The recent observational result of the cluster IC 348
\citep[][]{Lada2006} also showed a dependence between the disk
fraction and the mass and spectral type of stars. Direct
observational evidence of truncated disks is also found in nearby
massive clusters, such as Orion. \citet{Vicente2005} probed the
disk size distribution in the Trapezium cluster. However, they
found no correlation between disk size and the distance to the
massive star or between disk size and the mass of the star, as
theories predict \citep[e.g.][]{Storzer1999}. They suggest that
there might be various mechanisms of disk destruction happening so
that no obvious correlation is shown.

In order to investigate the relationship between massive stars and
low mass young stars, we plot the SED slope of YSO candidates vs.
radial distance from HAEBEs (Figure 8). The normal stars are also
ploted as comparison. If the massive star truncates the
circumstellar disks of its companions, stars closer to the massive
star might reveal themselves earlier than they would normally.
Therefore, these stars might look older than those further away,
so there might be a correlation between the youthfulness (SED
slope) and the distance to the massive stars. From Figure 8, we
can see that there is a young population near the HAEBE star
differing from the large-scale, uniform distribution, especially
for BD40, VV Ser, and BD46. However, there is no obvious trend
between the SED slope (youthfulness of young stars) and the
distance from massive star. This suggests that at the scale
of our study, the clustering of young stars dominates over the effect
of massive stars on the low-mass stars. Indeed, the range showing
the existence of proplyds in Orion is $\sim$ 0.3 pc
\citep[][]{Vicente2005}. Because Orion is a much more massive
cluster than our sample, any similiar effect from HAEBEs would
probably occur on scales much less than 0.3 pc. However, for our sample, within this range,
there are typically $<$ 5 YSO candidates because of the resolution
of Spitzer and the saturation of the HAEBEs, so it is unlikely
to detect any effect from massive stars.

\subsection{YSOs in Group vs. Isolation Within the Same Cloud}

Although there is no obvious trend between the SED slope and the
distance to HAEBE of the YSO candidates in groups, Figure 8 shows
a possible age dispersion of YSO candidates throughout the
molecular clouds, especially between inside and outside the group.
Indeed, from Figure 8, the YSO candidates in groups around HAEBEs
tend to have larger SED slopes than those outside. This means that
there is an enhancement of not only number but also the
youthfulness of YSOs within groups compared to those outside.
Actually, this is not unexpected, especially for low mass
clusters, which often show hierarchical structures within the
natal molecular cloud. Previous observations of the giant
molecular cloud, Lynds 1641, suggest that there exist not only
aggregates of mainly low mass protostars, but also a distributed
population of young stars throughout the whole cloud
\citep[e.g,][]{Strom1993,Allen1995}. \citet{Strom1993} found more
young stars are within the aggregates than distributed.
\citet{Allen1995} found stars within the aggregates have average
ages less than those distributed. In our sample, a similar picture
to Lynds 1641 is seen. \citet{Allen1995} suggested that this is
because there is a constant star forming rate within the
aggregates that constantly disperses forming stars outward, that
produce those isolated young stars. Another explanation for
isolated young stars outside the aggregates is that they formed
isolated within the cloud initially so that the observational
result is showing a primordial age dispersion throughout the
cloud.

However, these results may be biased by the extinction,
which would redden stars to make them seem younger, especially in the
regions near the group center. According to the extinction estimation,
extinction would affect the SED slopes
by 1 at most with A$_V \sim$ 10. This is still less than
the difference ($\sim$ 2) of SED slopes between the majority
of YSO candidates inside
and outside the group.
Therefore, we can conclude
the deviation of SED slopes due to extinction
is not significant so that
there is indeed a trend showing YSOs in the groups are younger.

As a group of young stars can be defined using the surface
number density ($R_n$ from Figure 7), it can alternatively be
defined by the aggregates on the age dispersion from Figure 8.
They do not necessary have to be the same as the spatial distribution
can be different from the age distribution of young stars. For
example, assuming a single star forming epoch, if $R_n$ is smaller
than the aggregates on the age dispersion, some stars formed
together in the beginning may have escaped out of the boundary of
clusters; on the contrary, if $R_n$ is larger than the aggregates
on the age dispersion, the impact from massive stars may have
dominated the inner region of clusters. Applying the estimated
$R_n$ into Figure 8 shows that the sizes of the groups around
HAEBEs based on Figure 8 are consistent with $R_n$. This
consistency may suggest that the escape of YSOs from the group is
not yet significant for our sample. In other words, the primordial
fragmentation of the molecular cloud may still dominate the young
stellar population outside the group in our sample.

Again, there is a group of YSO candidates around the HAEBE
younger than those outside the group, especially for \bdstar, NGC
7129, VY Mon, VV Ser, {$\rm BD+46\arcdeg~3471$}, and {$\rm
BD+46\arcdeg~3474$}. Possible explanations of this trend are:
there is a higher star forming rate and continuous star formation
within the group than outside the group
\citep[as suggested in the Lynds 1641 study,][]{Allen1995},
groups tend to form after
isolated low mass stars because they need more time to collect
enough material to initial the formation, or stars in group form
slower than isolation due to the competition of material with
group members that delay the collapse of clumps in the
beginning. Similar studies are needed for massive clusters or low
mass clusters for further investigation.

\section{Conclusions}

We present the NIR and MIR observations of eight embedded young
stellar groups around Herbig Ae/Be stars (3-20 M$\odot$ at age
$\sim$ a few Myrs) using the 2MASS point source catalog and the
Spitzer archive. By using published infrared photometry data to
obtain intrinsic colors of stars in the nearby cluster IC 348 and
comparing to Taurus, we conclude that the color-color diagram of J
- [3.6] vs. Ks - [4.5] is the most suitable tool to identify young
stars for our sample. We do not use colors of [5.8] and [8.0]
because of their intrinsic difficulties at nebulosity confusion
and sensitivity. Moreover, YSO loci can be derived because all
dereddened disk stars are located within narrow bands in the
diagrams of J - [3.6] vs. Ks - [4.5] and H - [3.6] vs. Ks - [4.5].
These loci are useful to estimate the extinction of young stars.

With identified YSO candidates, we can study the cluster environment and
cluster formation by examining their spatial distribution and
infrared properties (such as SED slopes), compared with the
position and mass of the HAEBEs. Some results are summarized as follows.

\begin{itemize}

\item Six of the sources show bright diffuse emission at [8.0],
which traces the dust illuminated by the HAEBEs. The other two are
the lowest mass groups, VV Ser and HD 97048, which may not be
energetic enough to illuminate the surroundings.

\item The HAEBEs are not always located at the center of the dust
emission. \bdstar is at the intersection of two branches of dust.
VY Mon and NGC 7129 are at the edge of the dust nebulae. Others
are located at the center of symmetric dust nebulae.

\item There are indeed young groups, containing 10-50 young stars,
around HAEBEs to form a relatively isolated system separated from
the other regions within the large-scale molecular cloud. The
HAEBEs are always near the center of young stars.

\item The groups around HAEBEs seem to be the sub-structure of a
large young population within the cloud. The sizes of groups based
on the spatial population of young stars are similar to
sub-cluster structures found in more massive clusters. 
Further comparison between
these small groups and massive clusters are needed.

\item There is no obvious trend between SED slopes and the distance
to the HAEBEs, that suggests that the effect of clustering on
young stars dominates over the effect of massive stars, at
the scale of our study.

\item Young stars in groups around HAEBEs tend to
show larger SED slopes than those outside the groups. It
suggests that young stars in groups are even younger than those outside.
Some possible explanations are:
(1) there is a higher star forming rate and a continuous star formation rate
within the group than outside the group
(2) massive stars form later than
low mass stars within the large-scale molecular cloud, or
(3) low mass stars in the group form slower than isolated ones
due to the effect of clustering or the massive stars.
Comparisons with more massive clusters are needed in the future.

\end{itemize}

\acknowledgements

We thank Robert Gruendl for helping the IRAC data analysis,
and You-Hua Chu and Rosie Chen for many valuable discussions.
We acknowledge support from the Laboratory for Astronomical Imaging at the
University of Illinois, NSF AST 0228953, and Spitzer Grant \# 1277965 provided
by NASA.
This work is based on archival data obtained with the
Spitzer Space Telescope, which is operated by the Jet Propulsion Laboratory,
California Institute of Technology under a contract with NASA.
This publication makes use of data products from
the Two Micron All Sky Survey, which is a joint project
of the University of Massachusetts and the Infrared Processing
and Analysis Center/California Institute of Technology,
funded by the National Aeronautics and Space Administration
and the National Science Foundation.
Also, this research used the NASA/ IPAC Infrared Science Archive,
which is operated by the Jet Propulsion Laboratory,
California Institute of Technology, under contract with
the National Aeronautics and Space Administration.

%\bibliographystyle{apj}
%\bibliography{herbig}

\clearpage

\begin{deluxetable}{lccllcccc}
\tablewidth{0pt}
\tablecaption{Stellar Paremeters of the Sources}
\tablehead{
\colhead{Target\tablenotemark{a}} & \colhead{Distance} 
& \colhead{Massive\tablenotemark{b}} & \colhead{RA.} 
& \colhead{Dec.} & \colhead{M} & \colhead{Spectral} & \colhead{Age} 
& \colhead{Refs.}\\
\colhead{} & \colhead{(kpc)} & \colhead{Star} & \colhead{(2000)} 
& \colhead{(2000)} &\colhead{(M$_{\odot}$)} & \colhead{Type} 
& \colhead{(Myrs)} & \colhead{}}
\startdata
BD40 & 1.00 & \bdstar & 20\,20\,28.25&+41\,21\,51.3& 13.0 & B2e & 0.06 & 1,2\\
                  & & V1686\,Cyg & 20\,20\,29.34&+41\,21\,27.6 & 4.5 & B5e & 0.6 & 1,2\\
                  & & V1318\,Cyg-S & 20\,20\,30.56&+41\,21\,25.4 & - & mid\,A-Fe & 0.6 & 1,2\\
\tableline
NGC\,\,7129 & 1.00 & $\rm BD+65\arcdeg~1637\ $ & 21\,42\,50.18 &+66\,06\,35.2& 9.2 & B3e & - &1\\
                 & & LkHa\,\,234 & 21\,43\,06.68 & +66\,06\,54.6 & 8.5 & B3e & 0.1 & 1\\
                 & & $\rm BD+65\arcdeg~1638\ $ & 21\,43& +66\,06 & - & B2 & - & 3\\
\tableline
MWC\,\,297 & 0.45 & MWC\,\,297 & 18\,27\,39.60 & -03\,49\,52.0 & 26.5 & O9e & 0.1-1.0 & 1,4 \\
\tableline
VY\,\,Mon & 0.80 & VY\,\,Mon & 06\,31\,06.94 & +10\,26\,05.0 & - & B8e & 0.1 & 4 \\
\tableline
VV\,\,Ser & 0.44 & VV\,\,Ser & 18\,28\,47.86 & +00\,08\,40.0 & 3.3 & B9e & 0.6 & 1,4 \\
\tableline
HD\,\,97048 & 0.16 & HD\,\,97048 & 11\,08\,03.32 & -77\,39\,17.5 & 3.4 & A0e & $>$ 5.0 & 1,4\\
\tableline
BD46 & 0.90 & $\rm BD+46\arcdeg~3471\ $ & 21\,52\,34.10 & +47\,13\,43.6 & 7.3 & A0e & 0.1-0.3 & 1,5\\
          & & $\rm BD+46\arcdeg~3474\ $ & 21\,53\,28.85 & +47\,15\,59.9 & - & B0 & 1.0 & 5\\
\tableline
V921\,\,Sco & 0.8 & V921\,\,Sco & 16\,59\,06.78 & -42\,42\,08.4 & - & B0e & 0.1-1.0 & 4 \\
 & & GAL343.49-00.03 & 16\,59\,23.30 & -42\,33\,55.0 & - & -& -& 6 \\
\enddata
\tablenotetext{a}{Name of the cluster system used here.}
\tablenotetext{b}{Massive stars that are included within the same cluster system
or within the same molecular cloud.}
\tablerefs{(1) \citet{Hillenbrand1992} (2) \citet{Hillenbrand1995} 
(3) \citet{Harvey1984}
(4) \citet{Habart2003} 
(5) \citet{Herbig2002}
(6) \citet{Caswell1987} }
\end{deluxetable}

\clearpage

\begin{deluxetable}{lccccc}
\tablewidth{0pt}
\tablecaption{Spitzer Archives of the IRAC Observations}
\tablehead{
\colhead{Target} & \colhead{Program ID.} & \colhead{Exp. Time} 
& \colhead{No. of Frames} & \colhead{FOV\tablenotemark{a}} & \colhead{Refs.}\\
\colhead{} & \colhead{} & \colhead{(Sec. / Frame)} &  
\colhead{} & ($\arcmin$) $\times$ ($\arcmin$) & \colhead{}}
\startdata
BD40 & 6 & 10.4 & 4 & 12.4 $\times$ 14.5 & 1\\
NGC\,\,7129 & 6 & 10.4 & 4 &12.4 $\times$ 14.0 & 1\\
MWC\,\,297 & 6 & 10.4 & 3 &12.6 $\times$ 14.5 & 1\\ 
VY\,\,Mon  & 6 & 10.4 & 4 &12.4 $\times$ 14.5 & 1\\
VV\,\,Ser & 174 & 10.4 & 2 &08.6 $\times$ 23.4 & 2\\ 
HD\,\,97048 &36 & 96.8 & 10 & 13.0 $\times$ 15.0 & 3\\
BD46 & 6 & 10.4 & 2 &26.4 $\times$ 28.0 & 1\\
V921\,\,Sco & 192 & 1.2 & 2 &19.5 $\times$ 37.8 & 4\\
\enddata
\tablenotetext{a}{Final field of view of the data set used for each source.}
\tablerefs{(1) GTO young embedded cluster survey by Fazio et al. 
(2) C2D legacy program by Evans et al. \citep[][]{Evans2003}
(3) GTO deep IRAC imaging for brown dwarfs by Fazio et al.
(4) GLIMPSE legacy program by Churchwell et al. \citep[][]{Benjamin2003}.}
\end{deluxetable}

\clearpage

\begin{deluxetable}{lcccccccc}
\tablewidth{0pt}
\tablecaption{Detection Limits of the Sources}
\tablehead{
\colhead{Target} & \colhead{J} & \colhead{H} & \colhead{Ks} 
&\colhead{[3.6]\tablenotemark{a}} & \colhead{[4.5]\tablenotemark{a}} 
& \colhead{[5.8]\tablenotemark{a}} & \colhead{[8.0]\tablenotemark{a}}
& \colhead{Spectral Type\tablenotemark{b}} }
\startdata
BD40 & 17.00 & 15.85 & 15.95 & 15.11 & 15.06 & 13.20 & 12.37 & M4\\
NGC\,\,7129 & 16.99 & 16.09 & 16.99 & 15.62 & 15.72 & 14.00 & 12.88 & M4\\
MWC\,\,297 & 16.99 & 16.13 & 15.51 & 14.68 & 14.60 & 13.59 & 12.34 & M5\\ 
VY\,\,Mon & 16.87 & 15.84 & 14.99 & 13.98 & 13.53 & 13.43 & 12.53 & M4\\
VV\,\,Ser & 16.97 & 16.06 & 16.66 & 15.21 & 15.16 & 13.24 & 12.32 &M6\\ 
HD\,\,97048 & 16.73 & 16.08 & 15.10 & 14.89 & 14.17 & 14.09 & 13.83 & M9\\
BD46 & 16.97 & 16.28 & 16.78 & 15.72 & 15.76 & 13.44 & 12.46 & M5\\
V921\,\,Sco & 16.00 & 15.34 & 14.95 & 13.31 & 12.73 & 11.11 & 10.29 & M2\\
\enddata
\tablenotetext{a}{The lower limits at magnitude when the uncertainty criterion is 0.05 mag.}
\tablenotetext{b}{The spectral type of the dimest detected YSO candidates,
estimated based on [4.5] from Figure 3.}
\end{deluxetable}

\clearpage

\begin{deluxetable}{lccccc}
\tablewidth{0pt}
\tablecaption{YSO Detection of the Sources}
\tablehead{
\colhead{Target} & \colhead{Num. of YSOs} & \colhead{Num. of YSOs\tablenotemark{a}} 
& \colhead{$R$\tablenotemark{b}} & \colhead{$R_d$} & \colhead{$R_n$} \\
\colhead{} & \colhead{(whole field)} & \colhead{(in groups)} 
& \colhead{(pc)} & \colhead{(pc)} 
& \colhead{(pc)}  }
\startdata
BD40 & 134 & 74 & 0.9 & 1.5\tablenotemark{c} & 0.6 \\
NGC\,\,7129 & 106 & 53 & 0.7 & 1.0 & 1.0 \\
MWC\,\,297 & 92 & 24 & 0.5 & 0.6 & 0.5 \\ 
VY\,\,Mon  & 42 & 26 & 0.5 & 0.8 & 0.75 \\
VV\,\,Ser & 148 & 22 & 0.5 & 0.4 & 0.5 \\ 
HD\,\,97048 & 16 & $\sim$ 10 & - & - & - \\
$\rm BD+46\arcdeg~3471\ $ & 500 & $\sim$ 10 & - & - & 1.3\tablenotemark{d} \\
$\rm BD+46\arcdeg~3474\ $ & 500 & 238 & 1.5 & 2.4 & 1.5 \\
V921\,\,Sco & 1271 & 33 & 0.6 & 0.7 & - \\
GAL343.49-00.03 & 1271 & 160 & 1.3 & 1.2 & - \\
\enddata
\tablenotetext{a}{Number of YSO candidates within $R_d$.}
\tablenotetext{b}{The radius calculated by 
$R=\sqrt{3}$ pc $(N/300)^{1/2}$ \citep[][]{Adams2006},
where $R$ is the radius and $N$ is the number of YSOs within the groups.}
\tablenotetext{c}{This value actually includes the two branch of
dense dust around this system, which is overestimated for
the group itself.}
\tablenotetext{d}{It actually includes two separated nearby 
small groups of YSO candidates.}
\end{deluxetable}

\clearpage

\begin{figure}[ht]
\begin{center}
\includegraphics[width=0.7\textwidth]{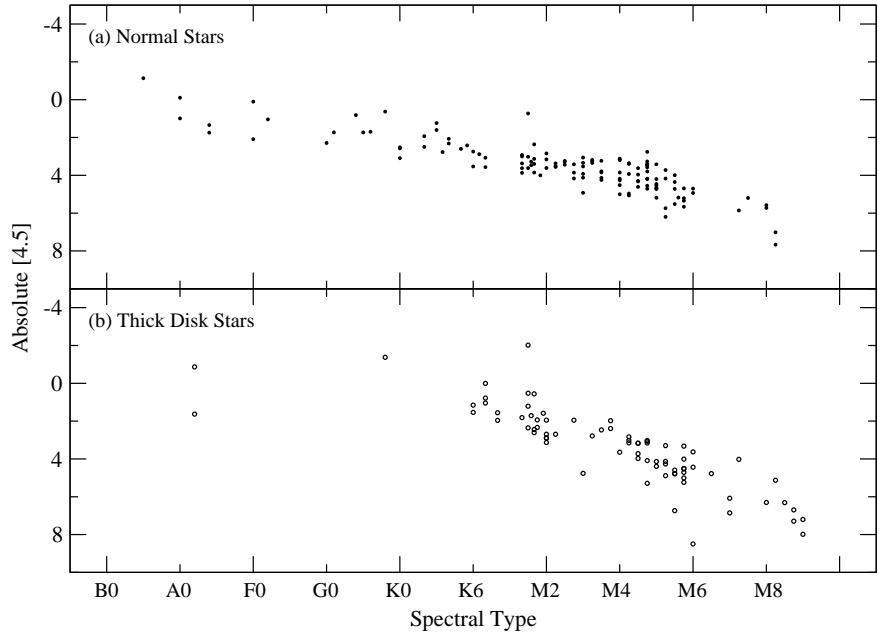}
\caption{The absolute magnitude at [4.5] vs. the spectral type
for normal (small dot) and thick disk (open circle) stars in IC 348
using the published data in \citet{Luhman2003} and
\citet{Lada2006}.}
\end{center}
\end{figure}

\begin{figure}[ht]
\begin{center}
\includegraphics[width=0.7\textwidth,angle=-90]{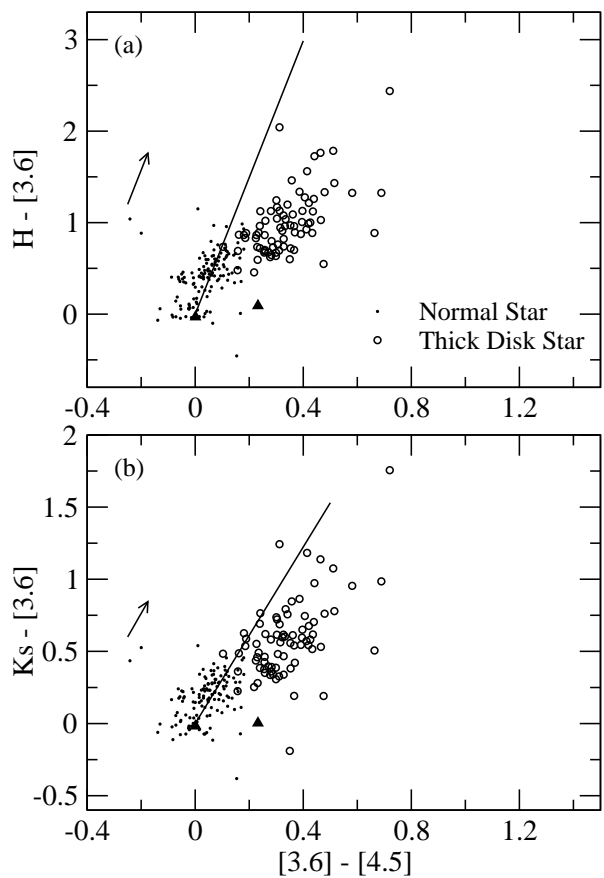}
\caption{The dereddened color-color diagrams of H - [3.6] vs.
[3.6] - [4.5] and Ks - [3.6] vs. [3.6] - [4.5] for normal and
thick disk stars in IC 348. The small dot and open circle label the
normal and thick disk stars (from G0 to M9), respectively. Two
solid triangles are A type thick disk stars (thus, Herbig Ae
stars). The solid lines are from the arrows that separate between
Class I/II and Class III protostars in Taurus in
\citet{Hartmann2005}. The small arrows on the left side are the
extinction vectors for A$_V$ $=$ 5.}
\end{center}
\end{figure}

\begin{figure}[ht]
\begin{center}
\includegraphics[width=0.7\textwidth,angle=-90]{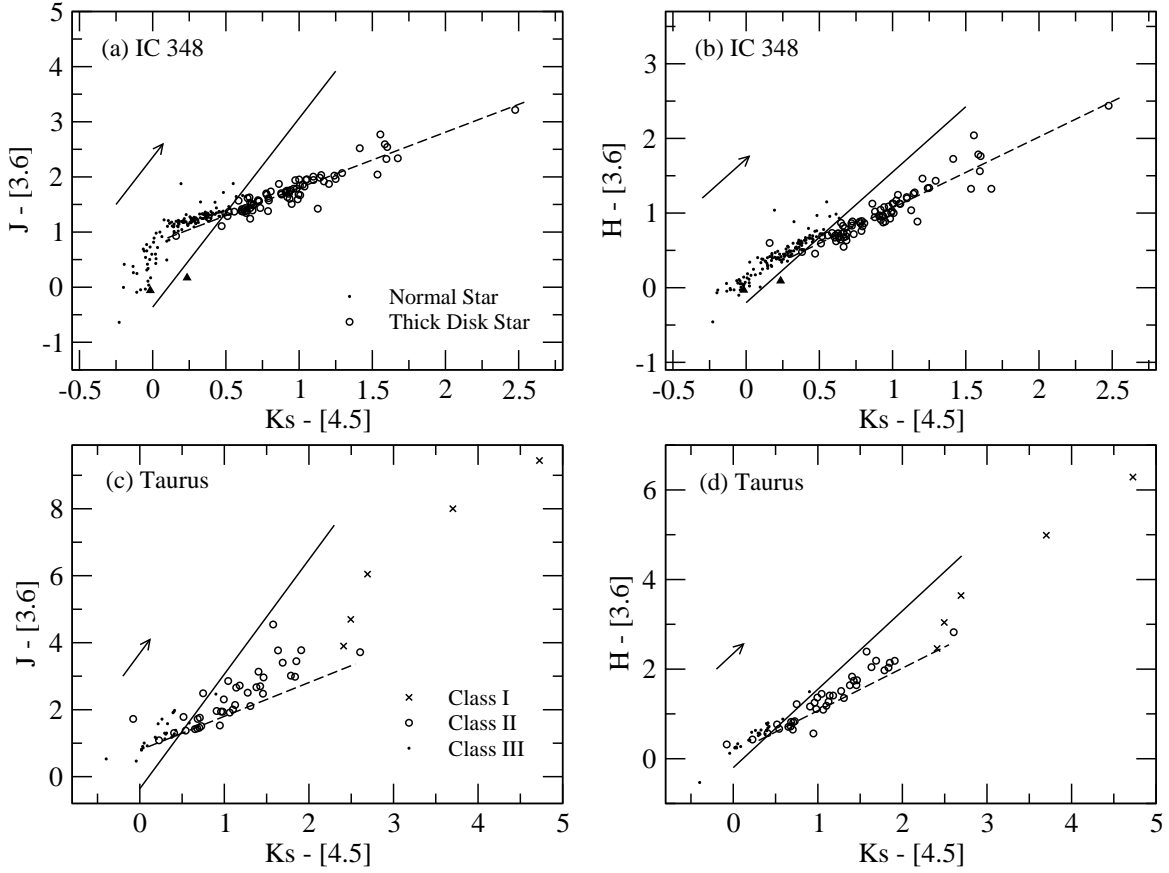}
\caption{(a) and (b) are the color-color diagrams of J - [3.6] vs.
Ks - [4.5] and H - [3.6] vs. Ks - [4.5] for dereddened normal and
thick disk stars in IC 348; while (c) and (d) are those (not
dereddened) for protostars in Taurus. In addition to the labels
described in the figure, the solid triangles are Herbig Ae stars.
The solid lines are Equation (1) for (a) and (c), and Equation (2)
for (b) and (d). The arrows show A$_V$ $=$ 5.}
\end{center}
\end{figure}

\clearpage
\begin{figure}[ht]
\begin{center}
\includegraphics[width=0.7\textwidth]{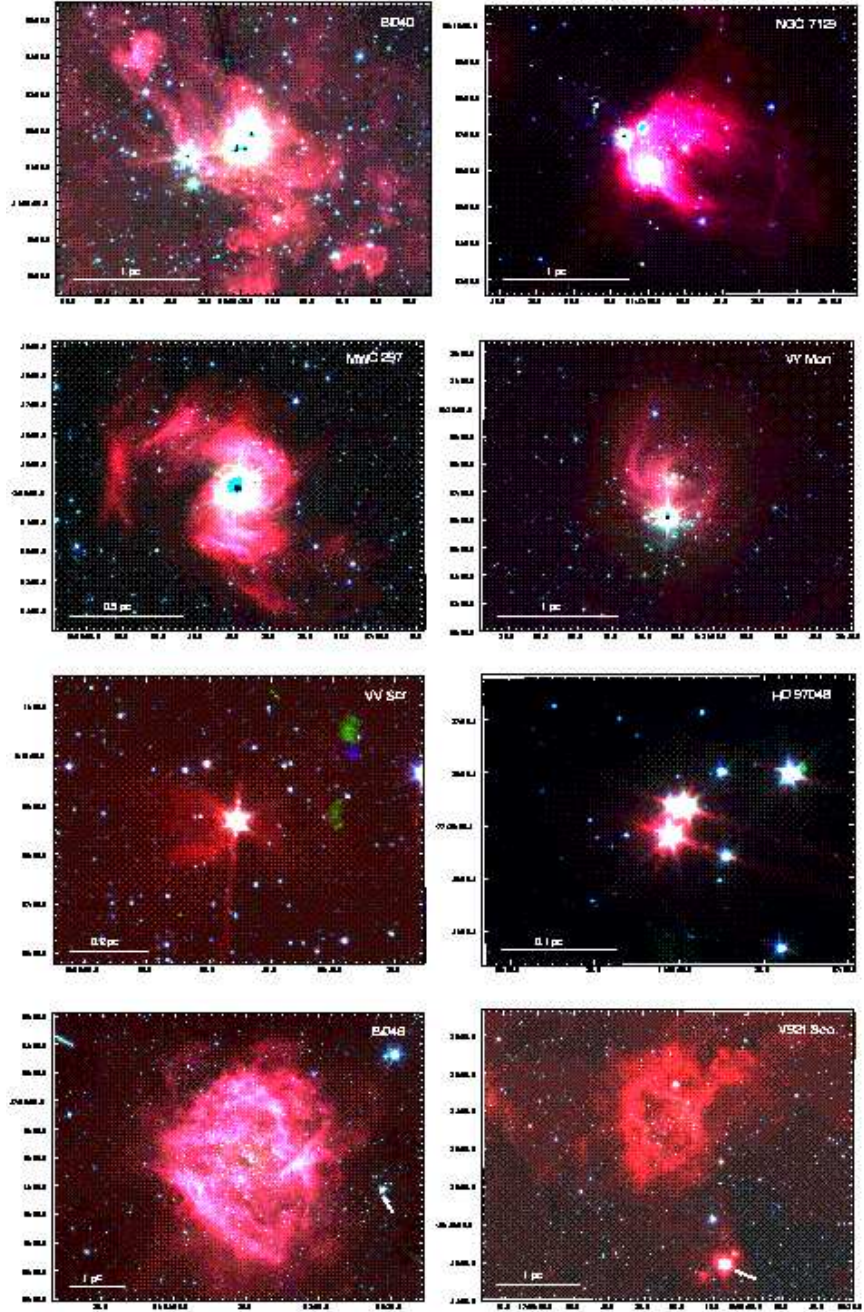}
\caption{Spitzer Images of our sample with coordinates in J2000. 
[3.6], [4.5], and [8.0] are
displayed as blue, green, and red, respectively. In the BD46
system, the B star {$\rm BD+46\arcdeg~3474$} is located at the
center of the bright nebula, while the Be star {$\rm
BD+46\arcdeg~3471$} is $\sim$ 3.5 pc away, as labeled by the
thick small arrow. Likewise, in the V921 Sco system, the big
bright nebula centers on an HII region, while the thick small
arrow points to the Be star V921 Sco at $\sim$ 2pc away.}
\end{center}
\end{figure}

\clearpage
\begin{figure}[ht]
\begin{center}
\includegraphics[width=0.9\textwidth]{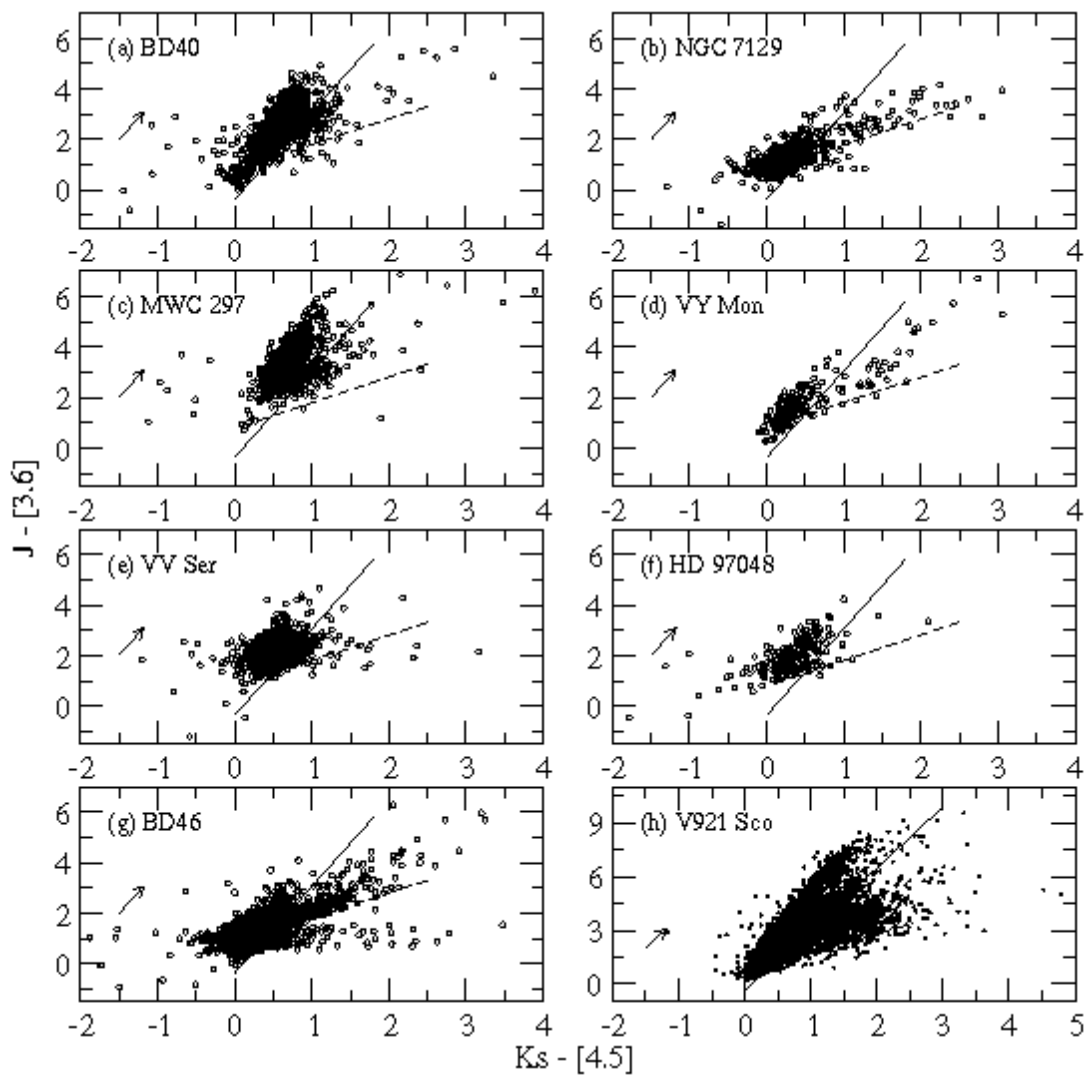}
\caption{The color-color diagrams of our sample. The solid lines are
Equation (1) and the dashed lines are the YSO loci (Equation (3)).
Stars that are located at the right side of the solid lines
are identified as YSO candidates.
The extinction vector for A$_V$ $=$ 5 is drawn as the arrow
in each panel.}
\end{center}
\end{figure}

\begin{figure}[ht]
\begin{center}
\includegraphics[width=0.7\textwidth,angle=-90]{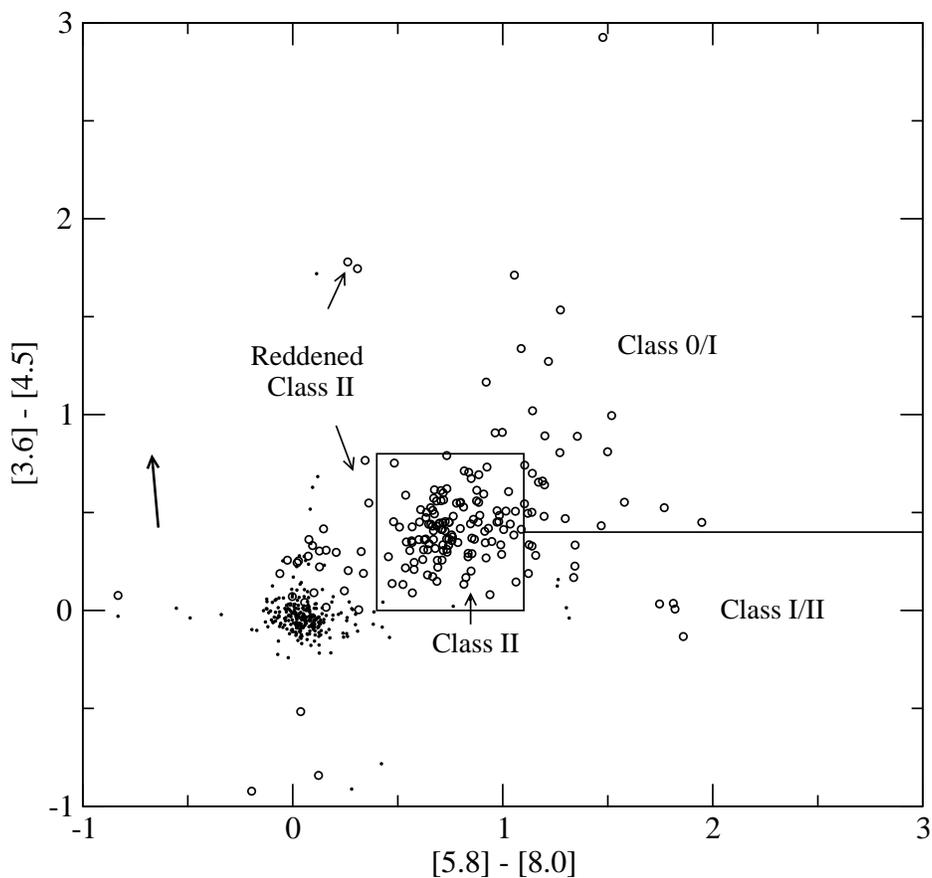}
\caption{The IRAC color-color diagram for those YSO candidates
with identifications at all four IRAC bands.
The thick vector on the left shows A$_V$ $=$30.
The small dot and circle label the normal stars and YSO candidates,
respectively. The labels for different types of protostars
are based on the classification in \citet{Megeath2004}.
Only 18 out of 322 (6 $\%$) YSO candidates are located
in the regime of normal stars.}
\end{center}
\end{figure}

\begin{figure}[ht]
\begin{center}
\includegraphics[width=0.8\textwidth,angle=-90]{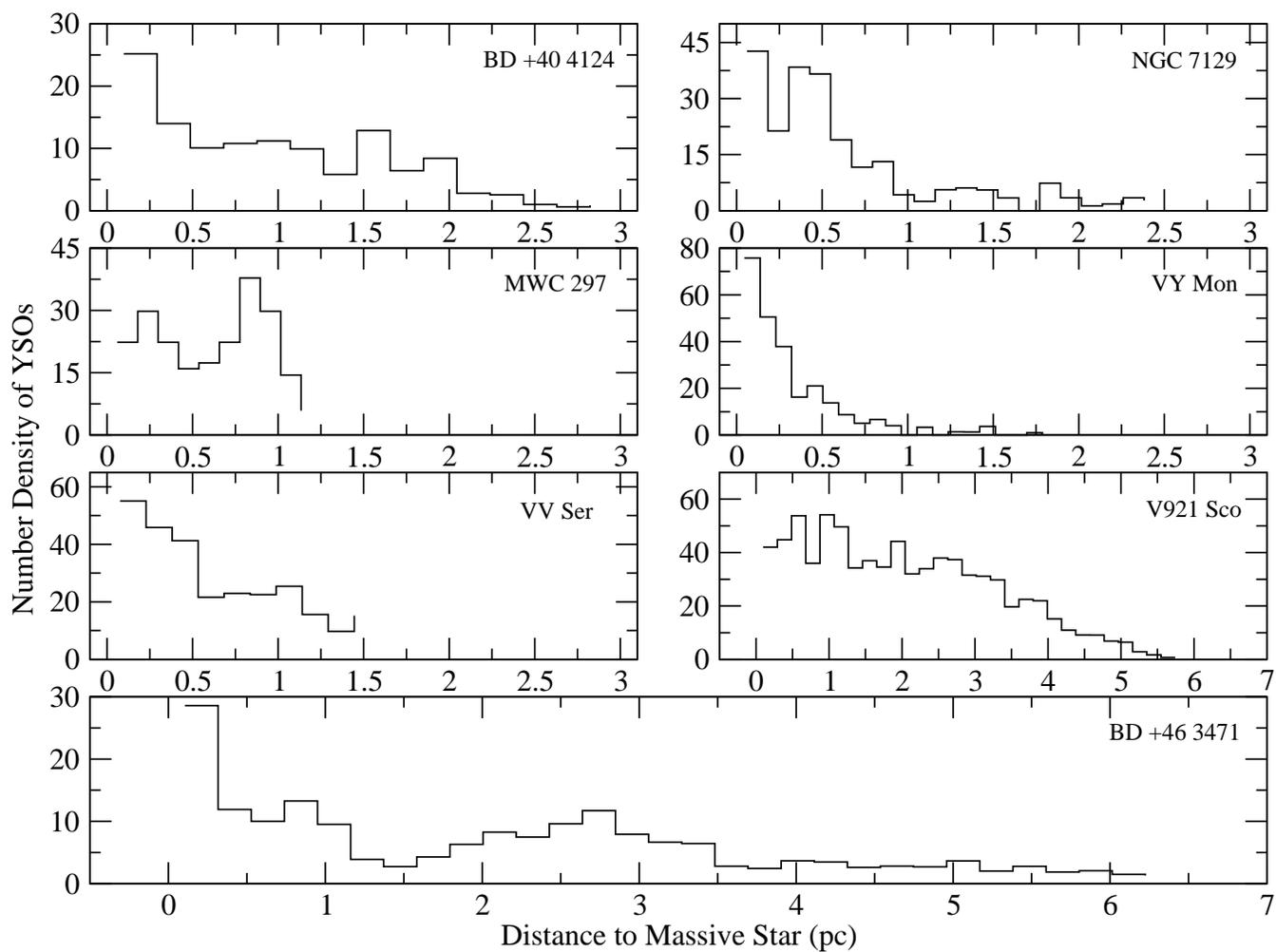}
\caption{The number density of YSO candidates vs. the distance to
massive stars shown at the up-right corner of each panel. See
texts in the Sec. 4 for the definition of the number density and
Table 1 for the position for each massive star. The group around
HD 97048 is not shown because there are too few stars to be
plotted.}
\end{center}
\end{figure}

\begin{figure}[ht]
\begin{center}
\includegraphics[width=0.8\textwidth,angle=-90]{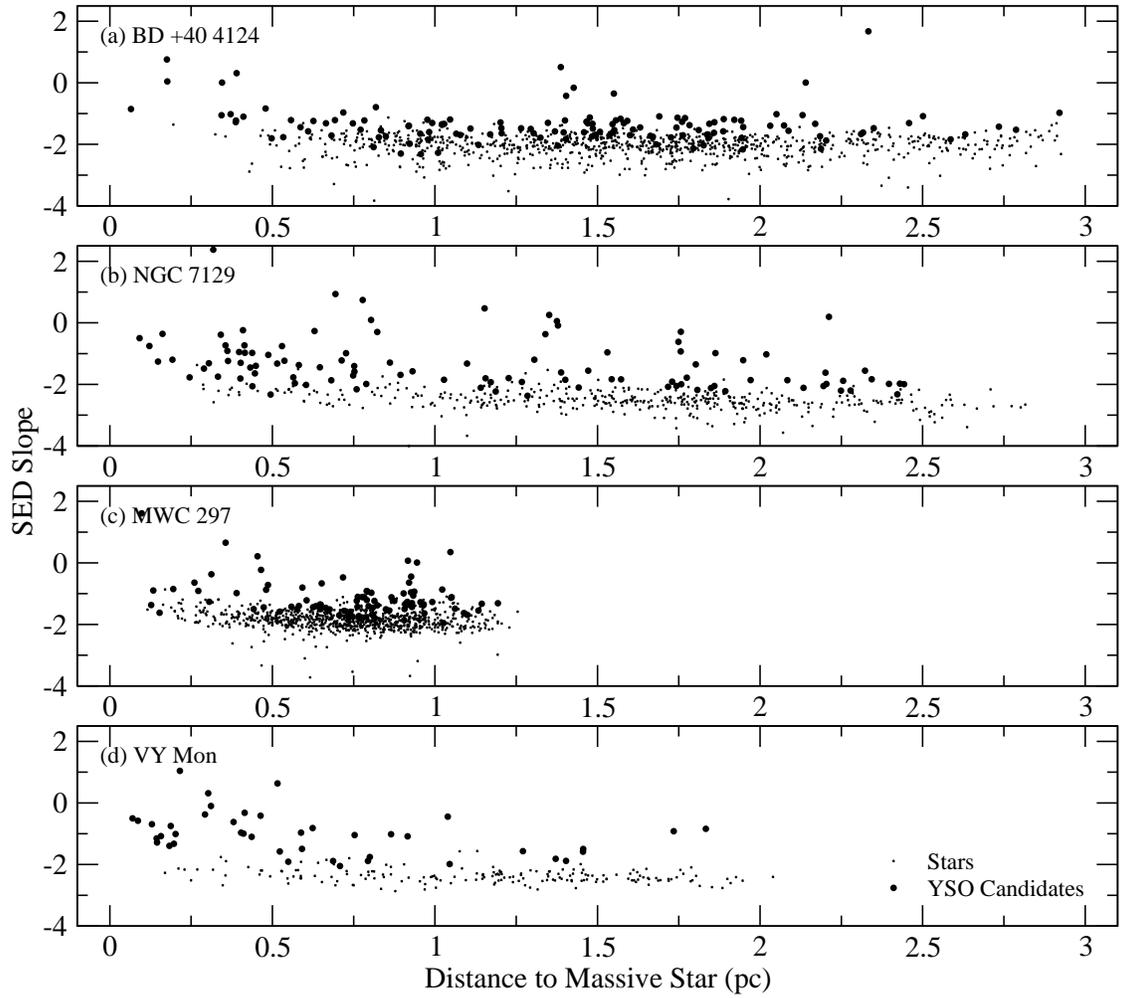}
\caption{The SED slopes of YSO candidates vs. the distance to
massive stars. The group of YSO candidates around the HAEBEs tend
to show larger SED slopes compared to those distributed outside the group.}
\end{center}
\end{figure}

\begin{figure}[ht]
\begin{center}
\includegraphics[width=0.9\textwidth]{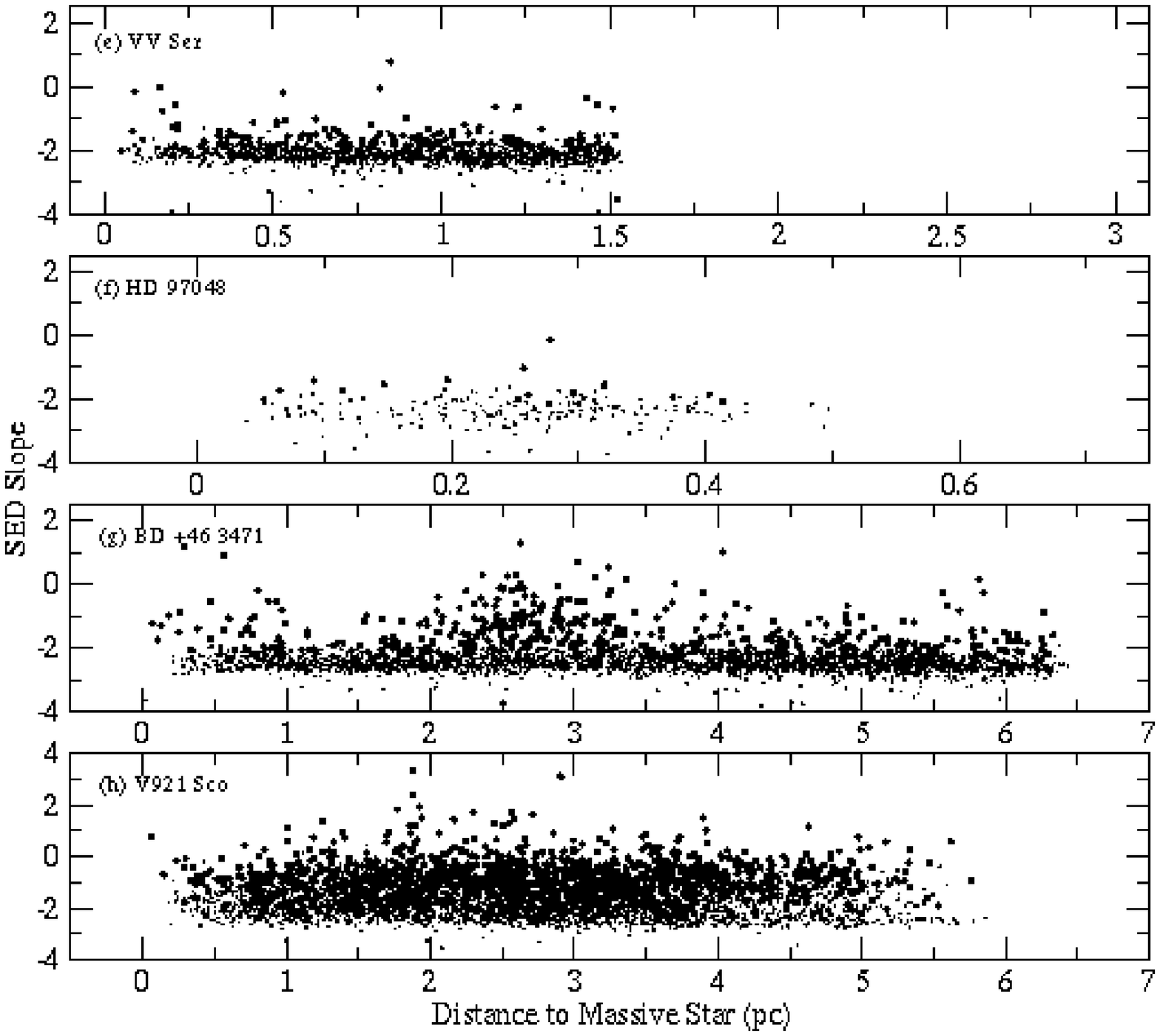}
\caption{Figure 8-- Cont.}
\end{center}
\end{figure}

\end{document}